\documentclass{aa501}
\usepackage{graphicx,amssymb}
\begin{document}

\title{Line-of-sight Effects on Observability of Kink and Sausage
Modes in Coronal Structures with Imaging Telescopes}

   \author{F.C. Cooper, V.M. Nakariakov, D. Tsiklauri}

   \offprints{Valery Nakariakov,\\ valery@astro.warwick.ac.uk}

   \institute{Physics Department,
   University of Warwick, Coventry, CV4 7AL, England.}

   \abstract{Kink modes of solar coronal structures, perturbing the loop in the
direction along the line-of-sight (LOS), can be observed as
emission intensity disturbances propagating along the loop
provided the angle between the LOS and the structure is not ninety
degrees. The  effect is based upon the change of the column
depth of the loop (along the LOS) by the wave. The observed
amplitude of the emission intensity variations can be larger than
the actual amplitude of the wave by a factor of two and
there is an optimal angle maximizing the observed amplitude.  
For other angles this
effect can also attenuate the observed wave amplitude. The
observed amplitude depends upon the ratio of the wave length of
kink perturbations to the width of the structure and on the angle
between the LOS and the axis of the structure.  Sausage modes are 
always affected negatively from the observational point of view,
as the observed amplitude is always less than the actual one.
This  effect
should be taken into account in the interpretation of wave
phenomena observed in the corona with space-borne and ground-based
imaging telescopes.  }


   \date{Received ???? 2002 / Accepted ???? 2002}

   \titlerunning{Observability of kink modes...}
\authorrunning{Cooper, Nakariakov \& Tsiklauri}

   \maketitle

\section{Introduction}

In last few years, significant progress in the observational study
of MHD wave activity of the solar corona has been achieved with
SOHO/EIT and TRACE EUV imaging telescopes. Flare-generated
decaying oscillations of coronal loops have been observed and
interpreted as kink fast magnetoacoustic modes of the loops
(Aschwanden et al. 1999; Nakariakov et al. 1999; Schrijver \&
Brown 2000; Aschwanden et al. 2002). Fast magnetoacoustic waves
are possibly responsible for events such as coronal Moreton (or
EIT) waves (Thompson et al. 1998, Ofman \& Thompson 2002). Slow
magnetoacoustic waves have been discovered in polar plumes (Ofman
et al. 1997; DeForest \& Gurman 1998; Ofman, Nakariakov \&
DeForest 1999) and in long loops (Berghmans \& Clette 1999; De
Moortel, Ireland, Walsh 2000; Nakariakov et al. 2000; De Moortel
2002 ). These observational breakthroughs give rise to the use of
MHD coronal seismology (Nakariakov et al. 1999; Robbrecht et al.
2001; Nakariakov \& Ofman 2001) and were interpreted to support
the idea of wave-based theories of coronal heating (e.g.,
Tsiklauri \& Nakariakov 2001), and the solar wind acceleration
(e.g., Ofman, Nakariakov \& Seghal 2000).

Slow and fast magnetoacoustic waves are compressive and cause
perturbations of plasma density. As an emission depends upon
the density, the waves can be detected as the emission variations by
imaging telescopes. An important characteristic of the phenomenon
is the angle between the direction of the wave propagation and the
line of sight (LOS). Imaging telescopes allow us to observe
magneto\-acoustic waves propagating at sufficiently large angles
to the LOS. In particular, this fact motivated the interpretation
of the propagating EUV emission disturbances as  slow
magnetoacoustic waves (see the references above). In addition,
Alfv\'en waves, which are linearly incompressible, as well as
almost incompressible kink modes of coronal magnetic structures
(e.g., Roberts 2000 and references therein), can also be detected
with an imaging telescope with sufficient spatial and
temporal resolution, if perturbations of the magnetic field have
a component {\it perpendicular} to the LOS. Indeed, as the
magnetic field is frozen into the coronal plasma, the
perpendicular displacement of the field can be highlighted by
variation of emission intensity.

In this paper we discuss an alternative method for the
observational detection of kink modes of coronal magnetic
structures oscillating in the plane {\it containing} the LOS. It
is shown that this would lead to modulation of the intensity of
the emission {\it along the axis of the structure}, produced by
the change of the LOS column depth of the loop. We also
demonstrate that this phenomenon is important for sausage modes.

\section{Kink modes of cylindric magnetic structures}

Kink modes of coronal loops, observed in particular, with the TRACE
EUV imaging telescope (see the references above), are periodic
transverse displacements of the magnetic flux tube forming the
loop. They should be distinguished from sausage modes which do not
perturb the tube axis. Modeling the loop tube as a straight
magnetic cylinder uniform along the axis, Edwin \& Roberts (1983)
found that the kink modes can be either surface or body, depending
upon the structure of the mode inside the tube. Also, the modes
can be slow or fast, corresponding to fast and slow
magnetoacoustic waves modified by the structuring of the medium.
In particular, in the low-$\beta$ plasma of the solar corona,
coronal loops can support fast and slow kink body modes.

In the case of a kink mode the loop tube oscillates almost as whole and
the cross-section of the loop is practically not perturbed by the
oscillation. Also, the density perturbation inside the loop is
insignificant in this mode. Indeed, for fast magnetoacoustic waves
in a low-$\beta$ coronal plasma, the field-aligned flows $V_z$ are
much smaller than the transverse motions $V_x$. Consequently, from
the continuity equation, one gets that the density perturbations
$\tilde{\rho}$ are connected with the transverse perturbations by
the expression
\begin{equation}
\frac{\tilde{\rho}}{\rho_0} \approx \sqrt{1 -C_{A0}^2k^2/\omega^2}
\ \frac{V_x}{C_{A0}}, \label{dens}
\end{equation}
where $\rho_0$ and $C_{A0}$ are the density and the Alfv\'en speed
inside the loop tube, respectively, and other notations are
standard. As the frequency $\omega$ of the fast kink mode of a
coronal loop tends to the value $C_{A0}k$ with the increase of the
wave number $k$ (Edwin \& Roberts 1983, Fig.~4), the coefficient
on the right hand side of Eq.~(\ref{dens}) tends to zero in the
same limit. Consequently, the fast kink mode is almost
incompressible. We would like to stress that the characteristic
spatial scale of the mode {\it inside} the loop tube is quite
different from the loop tube diameter. In the limit $\omega/k \to
C_{A0}$, the characteristic transverse scale inside the tube tends
to infinity, i.e., inside the tube diameter the density is
not perturbed. {\it Outside} the tube, the scale is different and
it determines the mode localization length, which is about
$\sqrt{\omega^2/C^2_{Ae}-k^2}$, where $C_{Ae}$ is the Alfv\'en
speed in the external medium. Consequently, outside the loop tube,
the kink mode can be quite compressible.

The fast kink mode is often confused with the true incompressible
Alfv\'en mode. However,  the Alfv\'en mode modified by
structuring is the torsional mode which, in the linear limit
and cylindrical geometry, does not perturb the tube
boundary.

Consider a kink mode which symmetrically perturbs the boundary of
the loop tube. The wavelength is assumed to be much shorter than
the loop length, so the loop can be considered as a straight
cylinder. The plane of the transverse oscillations produced by the
kink mode forms an angle $\alpha$ with the line-of-sight (LOS). In
the following, we assume that the angle $\alpha$ is zero, i.e.
LOS is co-planar with the plane of oscillation.
The tube axis has an angle $\theta$ with LOS (see Figure~1). We
neglect the perturbations of density by the mode, assuming that
the density is equal to the unperturbed density $\rho_0$. The tube
diameter is $w$. We also assume that the diameter of the
tube does not change in time and space. The intensity of the
emission $I$ produced by the loop is proportional to the LOS column
depth of the loop tube $l$ and to the density of the plasma squared,
\begin{equation}\label{inten}
  I \propto \rho_0^2 l.
\end{equation}
In the absence of the perturbations, the intensity $I$ along a
straight segment of the loop is constant, $I \propto \rho_0^2 w /
\sin\,\theta$.

When the loop experiences kink perturbations, the column
depth $l(d,t)$ of the loop tube depends upon the coordinate
$d$ along the loop (taking into account the effect of projection,
see Fig.~1) and changes with time $t$.
As, in the optically thin medium, the intensity of the emission is
proportional to thickness of the volume of plasma emitting the
radiation, the variation of the LOS column depth causes the
variation of the intensity.
Thus, the emission intensity variations can be produced even by
entirely {\it incompressible} kink waves, and consequently, by the
almost incompressible kink modes of coronal loops.

\begin{figure}
     \resizebox{\hsize}{!}
     {\includegraphics{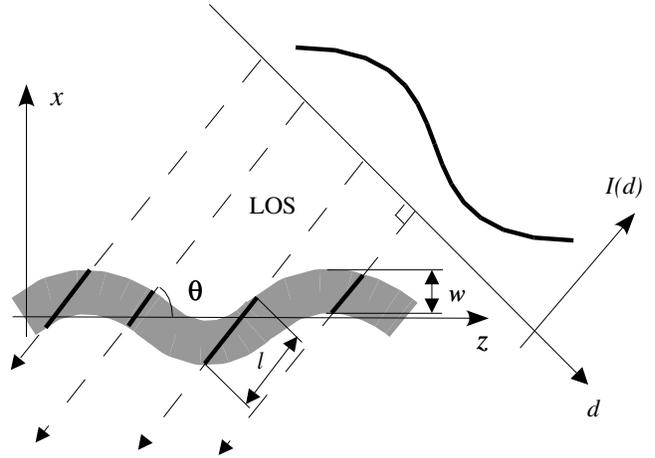}}
     \caption{A snapshot of perturbation of a segment of a coronal loop
     by a harmonic kink wave and produced variations of the emission intensity
     along the loop image. The variations produce
     the change of the LOS column depth of the loop if the angle $\theta$
     between the loop axis and the line-of-sight differs from ninety degrees.
     The variations of the column depth of the loop 
     modulate the emission intensity
     $I(d)$ on the loop image.}
     \label{fig1}
\end{figure}

Note that the density perturbations produced by the wave outside
the loop are ignored in Eq.~(\ref{inten}), as the external plasma
is observed to be much more rarefied.

\section{Parametric studies}

\subsection{Kink modes}

\begin{figure*}
\centering
 \includegraphics[width=17cm]{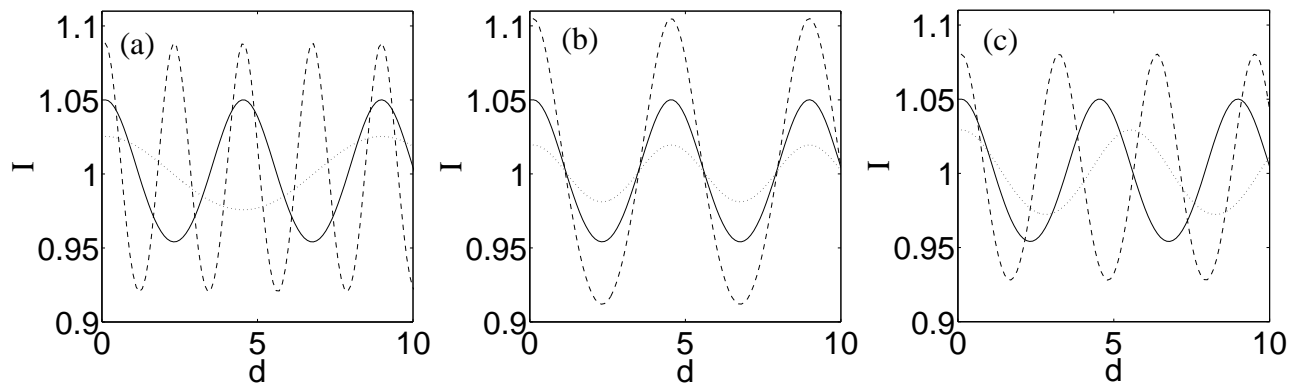}
\caption{ The effect of changing 
wavelength (a), wave amplitude (b) and viewing angle (c) respectively 
on the observed emission intensity variations along a straight
segment of a coronal loop in the presence of a harmonic kink wave.
The intensity $I$ is normalized to the unperturbed intensity in
the absence of the wave. The distance $d$ is normalized to the
loop diameter. On the left panel, the dotted curve
corresponds to the wavelength $\lambda = 4\pi$, the solid to
$\lambda = 2\pi$ and the dashed to $\lambda = \pi$, measured in
 loop diameters. For all  curves, the normalized wave
amplitude is $a=0.05$ and the angle between the loop axis and the
LOS is $\theta=\pi/4$. On the middle panel, the dotted curve
corresponds to $a=0.02$, the solid to $a=0.05$ and the dashed to
$a=0.1$, for $\lambda = 2\pi$ and $\theta = \pi/4$. On the right
panel, the dotted curve corresponds to $\theta = \pi/3$, the solid
to $\theta=\pi/4$ and the dashed to $\theta=\pi/6$, for $a=0.05$
and $\lambda = 2\pi$. } \label{snaps}
\end{figure*}
\begin{figure}
     \resizebox{\hsize}{!}
     {\includegraphics{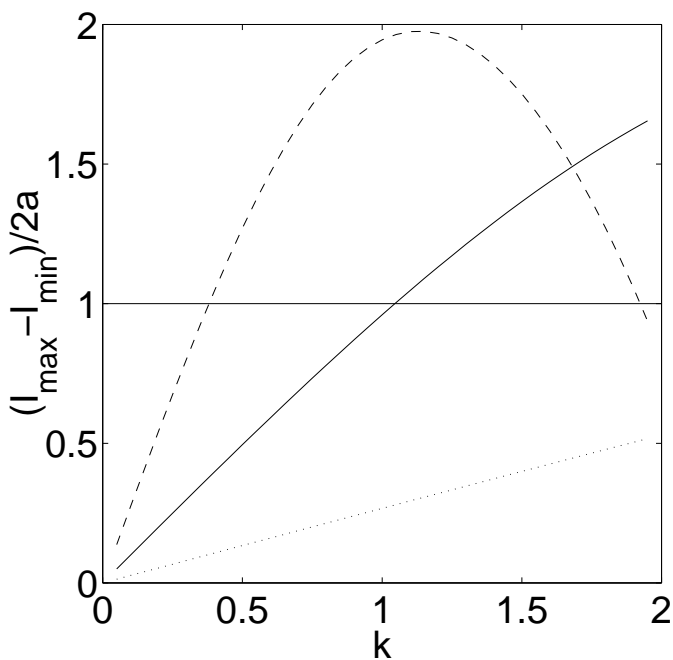}}
     \caption{Dependence of the observed amplitude of emission intensity
variations along a straight segment of a coronal loop in the
presence of a harmonic kink wave upon the wave number $k$,
normalized to the loop diameter  and the wave amplitude, which is 0.05
for all three curves. The dotted curve corresponds to the angle between
the loop axis and the LOS $\theta=5 \pi /12$ $(75\degr)$, the solid to
$\theta=\pi/4$ $(45\degr )$ and the dashed to $\theta=\pi/9$ $(20\degr )$.
The solid horizontal straight line indicates the actual normalized wave
amplitude. Points above this line indicate amplification while points
below indicate attenuation.}
     \label{on_k}
\end{figure}
\begin{figure}
     \resizebox{\hsize}{!}
     {\includegraphics{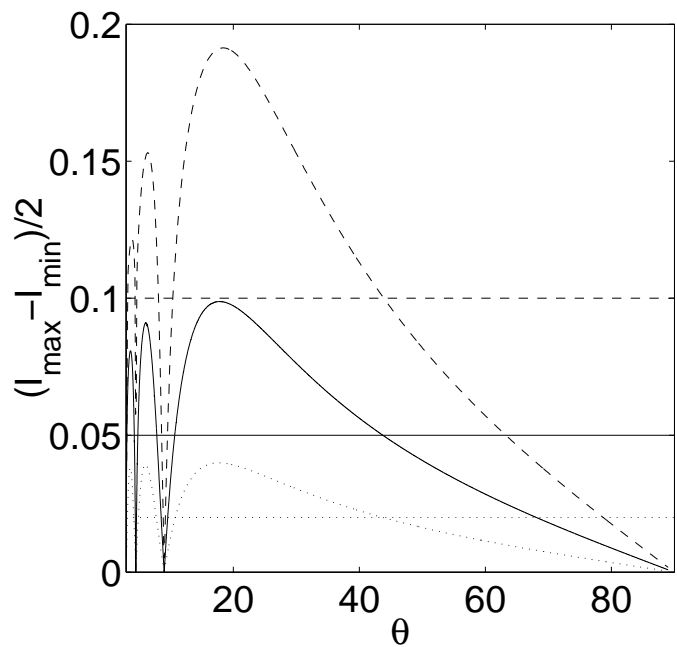}}
     \caption{Dependence of the observed amplitude of emission intensity
variations along a straight segment of a coronal loop in the
presence of a harmonic kink wave upon the angle between the loop
axis and the LOS $\theta$. The dotted curve corresponds to the
wave amplitude $a=0.02$, the solid to $a=0.05$ and the dashed to
$a=0.1$, for the normalized wavelength $\lambda=2\pi$.  Similarly the
straight horizontal lines represent the actual amplitude of their respective
waves. Points above these lines indicate amplification while points
below indicate attenuation.}
     \label{on_theta}
\end{figure}

Consider a snapshot of a harmonic kink perturbation of a straight
tube modeling a segment of a coronal loop. At a given time,
boundaries of the tube are given by the equations
\begin{equation}
x=a \sin \left( kz \right) \pm \frac{w}{2}, \label{kink:eqn}
\end{equation}
where $a$ is the perturbation amplitude and $k$ is the wave number
($k=2 \pi / \lambda$ where $\lambda$ is the wavelength), and the
plus and minus signs correspond to the upper and lower boundaries,
respectively. The geometry of the problem is shown in Fig.~1.
Observing the tube at an angle $\theta$, we see that the LOS
column depth of the tube is modulated by the kink perturbation.
The family of parallel LOS is given by the equation
\begin{equation}
x=z \tan \theta - d/\cos \theta, \label{sight:eqn}
\end{equation}
where the parameter $d$ represents the coordinate across the LOS
and, consequently, along the image (see Fig.~1).

Solving the set of equations (\ref{kink:eqn}) and
(\ref{sight:eqn}), we find the intersections of the LOS with the
tube boundaries. We then have two points, one where the LOS enters
the tube and one where it exits. Using these points we can determine
the column depth of the tube along a
given LOS as a function of the coordinate $d$ along the image.
Note that the line of sight may pass into and out of the flux tube
more than once for certain amplitudes, LOS angles and wave
numbers. We must therefore take into account all solutions of the
set (\ref{kink:eqn})-(\ref{sight:eqn}) and whether the line of sight is
inside or outside of the tube. The LOS passing through the tube more
than once only occurs if the gradient of the normal to the LOS is
greater than the minimum gradient of equations (\ref{kink:eqn}), i.e.
\begin{equation}
ak>\frac{1}{\tan \theta}.
\end{equation}
Thus, we determine the LOS column depth
of the tube as a function of the position on the image,
which, if the density of plasma inside the tube and the width of
the tube across the LOS remain constant, is proportional to the
intensity of the emission coming from the tube. The intensity
is normalized to the intensity given by a tube at the same angle to
the LOS but with zero amplitude waves, $a=0$.  This measures
the effect the presence of waves have on intensity.

Figures~2--4 demonstrate how the effect discussed depends upon the
parameters of the problem. Fig.~2 shows the distribution of the
emission intensity along the tube on an image for different
wavelengths and  amplitudes of the kink mode and the observation
angles $\theta$. Obviously, the observed variation of the emission
intensity is proportional to the amplitude of the wave and it decreases
with the increase of the wavelength. Indeed, in the limiting
case when the wavelength tends to infinity, the LOS column depth
of the tube does not change and the effect vanishes. The observed
amplitude is calculated as the difference
between the maximum and the minimum values of the observed
intensity, divided by two. The
dependence of the observed amplitude of the intensity
perturbations upon the wave number of the kink wave is shown in
Fig.~3. We gather from this graph that: For the  $75\degr $ dotted line
there is a decrease in the attenuation with increasing $k$ for
$0<k<2$; for the  $45\degr $ solid line, for $0<k \lesssim 1$,
there is a decrease in the attenuation with increasing $k$,
while for $1 \lesssim k<2$ there is an increase in the
amplification; for the  $20\degr $ dashed line, for
the range of $k$'s conisdered, there is an optimal wave number
corresponding to the maximum amplification. 
To graphical accuracy the $75\degr $
and $45\degr $ curves with amplitudes $a=0.02$ and $a=0.1$ are identical
to the $a=0.05$ curves and the $a=0.02$ and $a=0.1$, $20\degr$ curves
(not plotted) lie very close to the $a=0.05$ dashed curve.
Therefore non-linear amplitude effects within the parameter regimes
considered are small to observational accuracy and not considered
further here.

\begin{figure}
     \resizebox{\hsize}{!}
     {\includegraphics{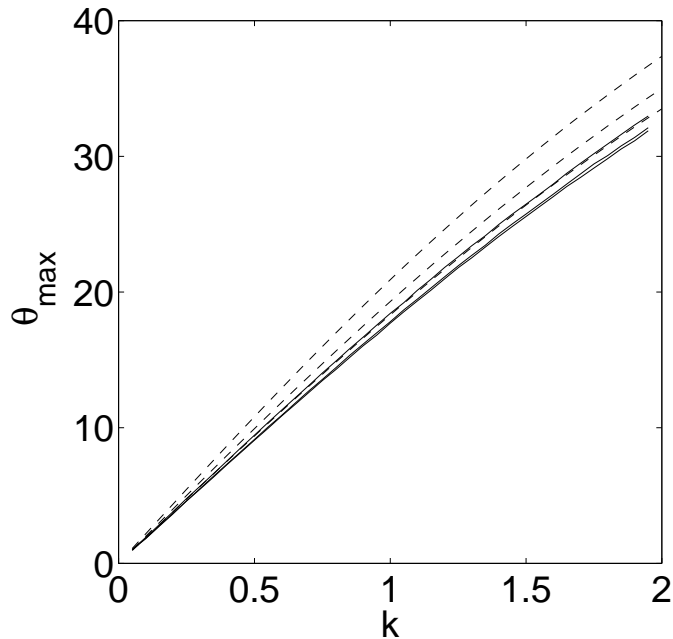}}
\caption{Dependence of the optimal observation angle along a straight
segment of a coronal loop in the presence of a harmonic kink wave upon
the wave number $k$. The solid curves corresponds to the actual optimal
angles and the dashed correspond to equation (\ref{tk:eqn}) for the
amplitude $a=0.1$, top, $a=0.05$, middle and $a=0.02$, bottom.
Note that the deviation between theory (Eq.(6)) and numerical
calculation is tolerably small, however, it grows with the increase of 
$a$ and $k$.}
\label{theta_k:fig}
\end{figure}

According to the right panel of Fig.~2 and Fig.~4, the dependence
of observed amplitude of the intensity variations upon the angle
$\theta$ has a pronounced maximum, corresponding to an optimal angle
of the wave detection. Indeed, the effect of the modulation of
the observed intensity by a kink wave vanishes when the angle
tends to $\pi/2$, while the whole approach fails when $\theta \to
0$ (the leftmost data points in Fig.~4 correspond to
$\theta=3^\circ$). Fig.~4 clearly demonstrates, however, that
there is an optimal angle which yields the maximum observed
amplitude. Thus, this phenomenon would make it possible to observe
kink modes in a certain segment of a coronal loop (in the
realistic curved geometry), where the angle $\theta$ is optimal.
This optimal angle varies with $k$ and weakly with $a$, figure
\ref{theta_k:fig}. It corresponds closely to a LOS passing through
a maximum of the upper curve and a neighboring  minimum
of the lower curve (see equations (\ref{kink:eqn})). Similarly,
there can be other optimal angles, if the LOS passes through a
maximum of the upper curve and another minima of the lower curve.
This estimation for the optimal angles $\theta_{max}$ gives
\begin{equation}
\theta_{max}=\arctan \left[ \frac{k \left( w+2a \right)}
{2 \pi \left( n-\frac{1}{2} \right)} \right],
\label{tk:eqn}
\end{equation}
where $n=1, 2, 3...$ is the peak number with decreasing $\theta$.
The highest  three rightmost peaks (for the same $\theta$, but different
amplitudes) in figure \ref{on_theta} 
correspond to $n=1$. Peaks corresponding to the higher values of $n=2$ and $3$
are  progressively 
smaller than the peaks
at $n=1$.  It is possible that the optimal angles smaller than this one
corresponding to $n=1$  are not relevant to coronal
applications, as for sufficiently small angles the curvature of
the loop plays an important role. The loop  may only be modeled
by a straight cylinder  if the approximately straight loop section
is sufficiently long.

There are also angles corresponding to zero intensity variation,
the first is at $\theta=90\degr $ (see Fig.~4). Here a LOS passes through the
maxima of both curves and the angles are given by
\begin{equation}
\theta_{min}=\arctan \left[ \frac{k w}{2 \pi n} \right].
\end{equation}
 When the angle is about these values the observed amplitude is attenuated
and at these zero points kink waves may not be detected at all. Note that
these points do not depend on amplitude.  In figure \ref{on_theta} the 
straight horizontal lines at the respective wave amplitudes divide the graph
into two regions; above, where the observed signal is amplified, and below, where
it is attenuated. If these curves are normalized to the amplitude, $a$, then to 
graphical accuracy they collapse on top of each other everywhere
except for the vicinity of the peaks where they are still close.

\subsection{Sausage modes}

\begin{figure*}
\centering
 \includegraphics[width=17cm]{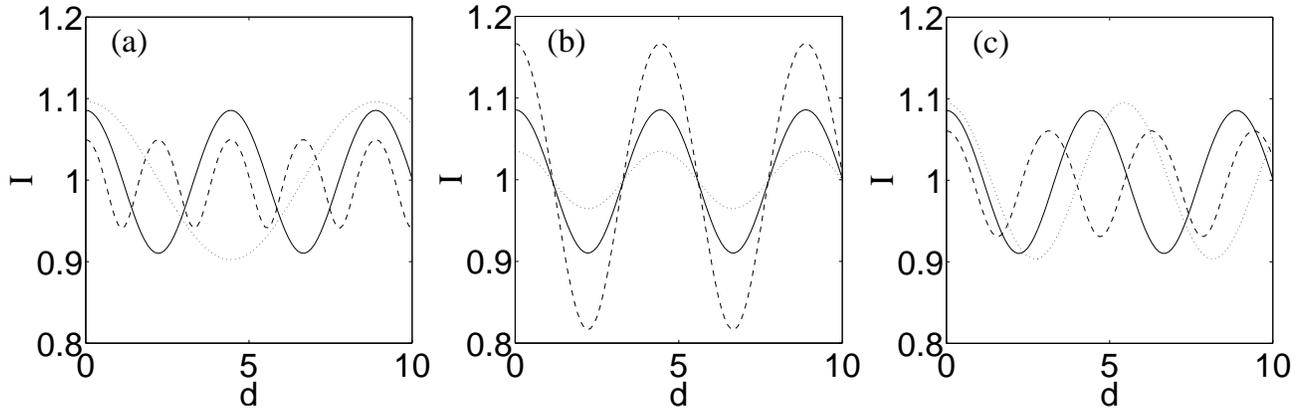}
\caption{ The effect of changing 
wavelength (a), wave amplitude (b) and viewing angle (c) respectively 
on the observed emission intensity variations along a straight
segment of a coronal loop in the presence of a harmonic sausage wave.
The parameters $\lambda$, $\theta$ and $a$ have identical
values to those given in figure \ref{snaps}.
} \label{snapsSaus}
\end{figure*}
\begin{figure}
     \resizebox{\hsize}{!}
     {\includegraphics{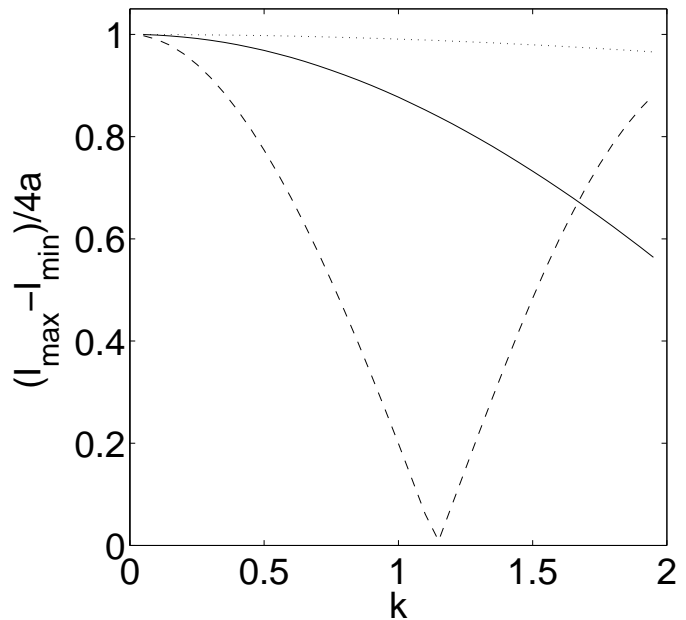}}
     \caption{Dependence of the observed amplitude of emission intensity
variations,  normalized to twice the wave amplitude, along a straight
segment of a coronal loop in the
presence of a harmonic sausage wave upon the wave number $k$,
normalized to the loop diameter.   $a=0.05$ for all three curves.
Curves of amplitudes $a=0.1$ and $a=0.02$
are almost identical to graphical accuracy. The dotted curve corresponds
to twice the angle between the loop axis and the LOS
$\theta=5 \pi /12$ $(75\degr )$, the solid to $\theta=\pi/4$ $(45\degr )$ and
the dashed to $\theta=\pi/9$ $(20\degr )$. Twice the normalized wave
amplitude is at $(I_{max}-I_{min})/4a=1$. All points fall
below this line, indicating attenuation of the expected measurements.
The minimum of the $20\degr $ curve is of a similar nature to the minimum
found in figures \ref{on_theta} and \ref{on_thetaSaus}.}
     \label{on_kSaus}
\end{figure}
\begin{figure}
     \resizebox{\hsize}{!}
     {\includegraphics{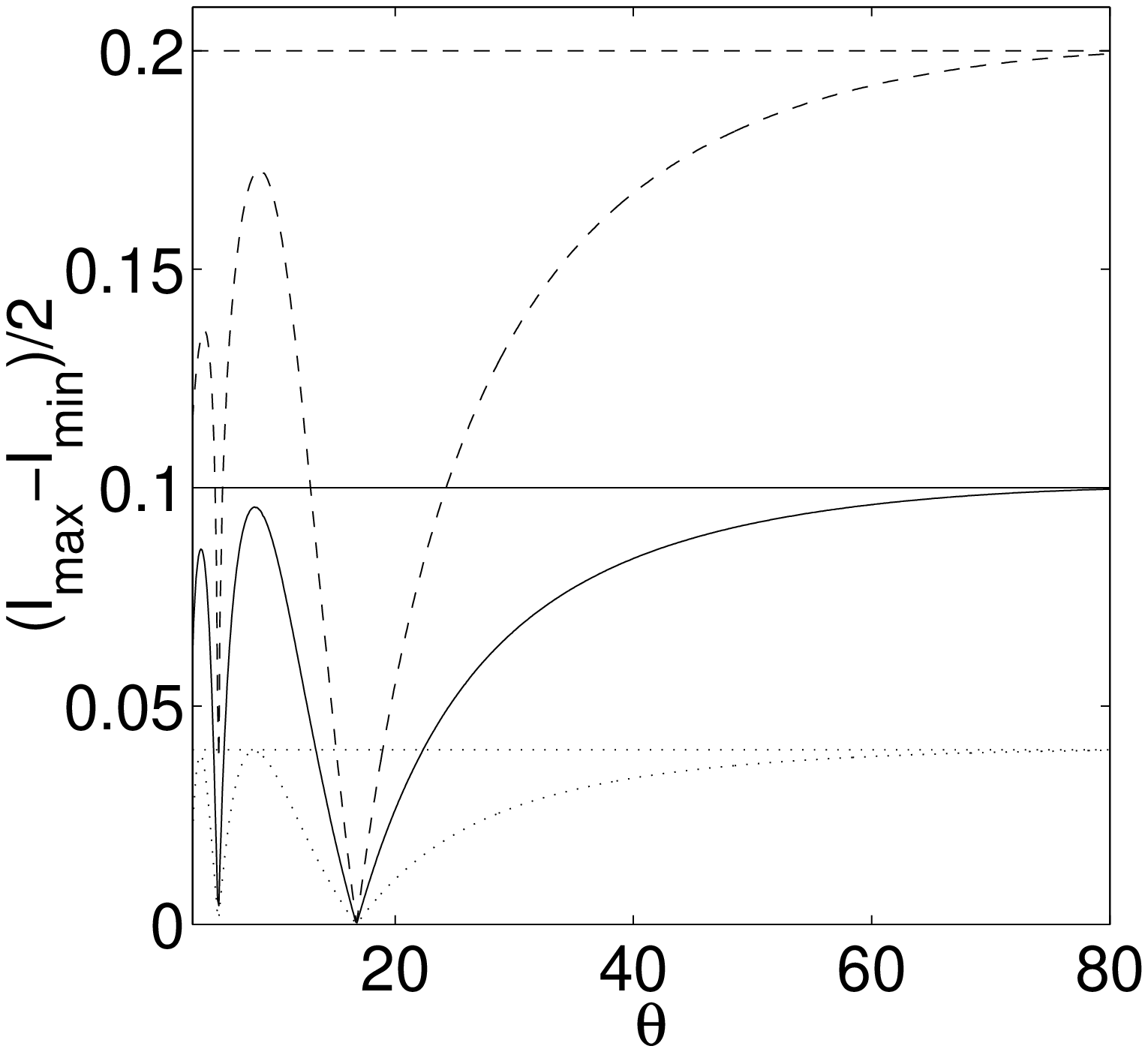}}
     \caption{Dependence of the observed amplitude of emission intensity
variations along a straight segment of a coronal loop in the
presence of a harmonic sausage wave upon the angle between the loop
axis and the LOS $\theta$. The dotted curve corresponds to the
wave amplitude $a=0.02$, the solid to $a=0.05$ and the dashed to
$a=0.1$, for the normalized wavelength $\lambda=2\pi$.  The
horizontal straight lines are added for comparison with figure \ref{on_theta}
and correspond to the amplitude of the respective
waves. Note that the amplitude
at $90\degr $ gives the largest amplitude and therefore we only see attenuation
below this.}
     \label{on_thetaSaus}
\end{figure}

Similarly, the effect discussed above should be important
for sausage modes too. Indeed, as the sausage modes are the
propagating perturbations of the loop tube cross-section, the LOS
column depth of the loop is modulated by these waves too. To
complete the picture, we now perform  a study of the LOS effect on
observability of sausage waves.

The boundaries of a straight tube exhibiting sausage oscillations
can be given by the equations
\begin{equation}
x=a \sin \left( kz+\frac{\pi}{2} \right) +\frac{w}{2}
\label{saussageTop:eqn}
\end{equation}
corresponding to the upper boundary and
\begin{equation}
x=a \sin \left( kz+\frac{3 \pi}{2} \right) -\frac{w}{2}
\label{saussageBottom:eqn}
\end{equation}
corresponding to the lower boundary. The family of parallel LOS is
given by equation (4). We solve these equations in
the same way as for the kink oscillations. Figure \ref{snapsSaus}
shows the distribution of emission intensity along the tube for
the same parameters as those given in figure \ref{snaps}. The
wavelength, $k$, and angle, $\theta$, at high angles
have the opposite effect of
kink oscillations on the amplitude intensity, figures
\ref{on_kSaus} and \ref{on_thetaSaus}. The best observation angle
is $\theta=90\degr $ as expected. Note that the analysis here is 
two dimensional and the observed intensity amplitude
would be increased for a three dimensional tube if the telescope
pixel size is of the order of the tube width. However only the
intensity scale of figures \ref{snapsSaus} to \ref{on_thetaSaus}
would change and not the form. The optimum angle for the
detection of sausage modes is still $90\degr $. The LOS only passes into
and out of the tube once for the angles considered.

\section{Conclusions}

The phenomenon of the modulation of the emission intensity by kink
modes polarized in the plane formed by the loop axis and the LOS
provides  a possibility for the observational detection of the
kink modes in coronal structures. According to the discussion
above, the LOS effect {\it at an optimal angle, $\theta_{max}$,} 
can amplify the kink perturbations by a
factor of 2 (cf.  Fig.~4). 
For example, if the boundary perturbation is produced
by a kink mode of a relatively modest amplitude of about 5\%,
which corresponds to the typical coronal wave amplitudes
detected by SOHO/EIT (DeForest \& Gurman 1998) and by TRACE (e.g.
De Moortel et al. 2002 and references therein), the observed
perturbation of the intensity produced by the kink wave can reach
10\%, which would make the wave easily observable. 
In contrast, the optimal observation angle for 
the sausage modes is simply $90\degr $ (cf.  Fig.~8).
 Consulting figures \ref{on_k} and \ref{on_theta}
we can see parameter regimes for both amplification and attenuation
of the kink mode observations. Figures 
\ref{on_kSaus} and \ref{on_thetaSaus} demonstrate that relative
to $90\degr $ observations there is only attenuation of sausage
modes.
Additional
observability constraints are connected with the wave period and
length.  In the case of EUV imaging coronal telescopes, such as
EIT and TRACE, the observability of the waves is limited by the
telescope time resolution. For example, taking the kink speed
inside a loop to be 1000~km/s, which corresponds to the
estimations in Nakariakov et al. 1999, Nakariakov \& Ofman 2001,
the time resolution of about 30~s does not allow us to observe
wavelengths shorter than 30~Mm.

In the case of ground-based observations, when the loop is
observed in the green line bandpass, the observability is limited
by the spatial resolution of the telescope, which is usually over
a few arcsec, as the time resolution of such observations is
usually less then 1 sec (e.g, the cadence time of SECIS is
2.25$\times10^{-2}$~s and the pixel size is 4.07~arcsec, Williams
et al. 2001). In particular, this effect can be responsible for
propagating 6~second disturbances of the Fe\,{\sc xiv} green line,
discovered by the stroboscopic method in a solar eclipse data and
interpreted as fast magnetoacoustic waves (Williams et al. 2002,
see also Williams et al. 2001). Indeed, estimating the
wavelength of the perturbations at about 18~Mm (for the wave
period of about 6~s and the propagation speed of about 2~Mm/s), we
conclude that the waves are detectable with the time and spatial
resolution of the telescope. For loop widths of about 3-6~Mm, the
normalized wavelength is about 3-6, which gives the wavenumber $k$
of about 1-2. According to Fig.~3, the LOS effect discussed in
this paper could amplify the observed amplitude of the wave,
optimizing the observability of the modes.  According to
figure \ref{theta_k:fig}, for the parameters discussed, the optimal angle
would be about $20\degr -40\degr $. This makes the
phenomenon discussed here relevant to the interpretation of
Williams et al. (2002)'s results. However,  the confident
interpretation of the propagating disturbances in terms of the
kink waves requires detailed comparison of the observed results
and theoretical predictions. Also, interpretation
of these observations in terms of propagating kink modes should be
tested against another possible interpretation of the waves as
sausage modes.  The study of the possible relevance of the LOS
effect to the interpretation of short wavelength propagating waves
observed in the corona by Williams et al. (2002) is now in progress
and will be presented elsewhere.

 Also, this phenomenon should be taken into
account in the analysis of other examples of coronal wave
activity, in particular the slow waves in loops and polar plumes,
discussed in Introduction. However, direct application of the
results presented in this paper to the coronal slow
magnetoacoustic waves is not possible as the slow waves are
essentially compressible, and the density variation should be
taken into account in Eq.~(\ref{inten}). This phenomenon should
also be taken into account in interpretation of intensity
oscillations observed in prominence fine structures (e.g. Joarder,
Nakariakov \& Roberts 1997; Diaz et al. 2001). This suggests
another possible  development of this study.

\begin{acknowledgements}
FCC and DT acknowledge financial support from PPARC.
The authors are grateful to M.J. Aschwanden, L. Ofman,
B. Roberts and the anonymous referee for valuable comments.

\end{acknowledgements}

\end{document}